\documentstyle[aps,prc,epsfig,preprint]{revtex}
\begin{document}
\draft
\title{Search for the electric dipole excitations to the $3s_{1/2}
\otimes [2^{+}_{1} \otimes 3^{-}_{1}]$ multiplet in $^{117}$Sn}
\author{J. Bryssinck$^{1}$, L. Govor$^{2}$, V. Yu. Ponomarev$^{1}$
\thanks{Permanent address : Bogoliubov Laboratory of
Theoretical Physics, Joint Institute of Nuclear Research, Dubna, Russia},
F. Bauwens$^{1}$, O. Beck$^{3}$, D. Belic$^{3}$, P. von
Brentano$^{4}$, D. De Frenne$^{1}$, C. Fransen$^{4}$,
R.-D. Herzberg$^{4}$\thanks{Present address:
Oliver Lodge Laboratory, University of
Liverpool, Oxford Street, Liverpool, L69 7ZE, UK}, E. Jacobs$^{1}$,
U. Kneissl$^{3}$, H. Maser$^{3}$,
A. Nord$^{3}$, N. Pietralla$^{4}$, H.H. Pitz$^{3}$,
V. Werner$^{4}$}    
\address{$^{1}$Vakgroep Subatomaire en Stralingsfysica, Universiteit
Gent, Proeftuinstraat 86, 9000 Gent, Belgium}
\address{$^{2}$Russian Scientific Centre ``Kurchatov Institute'',
Moscow, Russia}
\address{$^{3}$Institut f\"{u}r Strahlenphysik, Universit\"{a}t Stuttgart,
Stuttgart, Germany} 
\address{$^{4}$ Institut f\"{u}r Kernphysik, Universit\"{a}t zu K\"{o}ln,
K\"{o}ln, Germany.}
\maketitle
\begin{abstract}
The odd-mass $^{117}$Sn nucleus was investigated in nuclear resonance
fluorescence experiments up to an endpoint energy of the incident
photon spectrum of 4.1~MeV at the bremsstrahlung facility of the
Stuttgart University.  More than 50 mainly hitherto unknown levels
were found.  From the measurement of the scattering cross sections
model independent absolute electric dipole excitation strengths were
extracted.  The measured angular distributions suggested the spins of
11 excited levels. Quasi-particle phonon model calculations including
a complete configuration space were performed for the first time for a
heavy odd-mass spherical nucleus. These calculations give a clear
insight in the fragmentation and distribution of the $E1$, $M1$, and
$E2$ excitation strength in the low energy region. It is proven that
the $1^{-}$ component of the two-phonon $[2^{+}_{1} \otimes
3^{-}_{1}]$ quintuplet built on top of the $1/2^{+}$ ground state is
strongly fragmented. The theoretical calculations are consistent with
the experimental data.  
\end{abstract}

\narrowtext

\section{Introduction}

Low-lying electric dipole excitations have been studied extensively in
a variety of spherical \cite{Knei96} and deformed nuclei
\cite{Zilg91,Knei93} over the last decade.  A survey on the
systematics of observed electric dipole excitations in the A =
130--200 mass region is given in Ref.~\cite{Fran98}.
Systematic nuclear resonance fluorescence (NRF) experiments performed
within 
the chains of the N = 82 isotones \cite{Herz95b} ($^{138}$Ba,
$^{140}$Ce, $^{142}$Nd and $^{144}$Sm) and the Z = 50 isotopes
\cite{Gova94b,Brys99} ($^{116-124}$Sn) showed that the low-lying electric
dipole strength $B(E$1)$\uparrow$ is mainly concentrated in the first
$J^{\pi} = 1^{-}_{1}$ state.  Some uniform properties of these
$1^{-}_{1}$ states were 
observed in both chains.  Their excitation energies are lying close to
the summed energies of the first quadrupole and octupole 
collective vibrational states and they are populated by an enhanced
electric dipole excitation (two to three orders of magnitude larger
than other low-lying $1^{-}$ states).  Both arguments, strongly
suggest an underlying quadrupole-octupole coupled $[2^{+}_{1}\otimes
3^{-}_{1}]$ two-phonon structure. Indeed, a detailed microscopic study
within the framework of the Quasi-particle phonon model revealed a
practically pure two-phonon $[2^{+}_{1}\otimes 3^{-}_{1}]_{1^{-}}$
configuration in the wave function of these $1^{-}_{1}$ states.  The
observed enhanced electric dipole strength of the ``forbidden'' $E1$
transitions can be reproduced from a consideration of the internal
fermion structure of the phonons and taking into account a delicate
destructive interference with the GDR $1^{-}$ one-phonons
\cite{Brys99,Pono98a}.  The most direct experimental proof for an 
underlying two-phonon structure can be obtained from a measurement of
the decay pattern of the two-phonon states to their one-phonon
components.  This has been achieved up to now only in very few cases:
$^{142}$Nd, $^{144}$Nd and $^{144}$Sm
\cite{Gate90,Robi94,Belg95,Wilh96,Wilh98}.  In each of these nuclei,
an enhanced $B(E2)$ strength in the decay of $1^{-}_{1} \rightarrow
3^{-}_{1}$ could be measured consistent with a two-phonon picture.  It
is still a tough challenge to observe the other members of the
$[2^{+}_{1} \otimes 3^{-}_{1}]$ two-phonon quintuplet as illustrated
recently in Ref.\ \cite{Garr99}.

As a natural extension of the systematic investigations on the
even-even spherical nuclei, the question arises how the observed
enhanced electric dipole excitation strength of the two-phonon
$[2^{+}_{1} \otimes 3^{-}_{1}]_{1^{-}}$ state fragments over several
levels of a particle two-phonon coupled multiplet in the odd-mass adjacent
nuclei. In such a study, the experimental technique and the
theoretical model should meet some requirements.  In the first place, the
experimental probe should be very selective in the excitation of
levels because of the high level density in odd-mass nuclei.  The
number of the involved levels in the odd-mass nucleus increases drastically
compared to the even-even neighbouring nucleus: in the odd-mass
nucleus levels with a spin equal $J_{0} - 1$, $J_{0}$ and $J_{0} 
+ 1$ can be populated via dipole excitations from the ground state
with spin $J_{0}$. Quadrupole transitions will excite
levels with a spin between $J_{0} - 2$ and $J_{0} + 2$. The real
photon probe is such a selective experimental tool.  Using an
intensive bremsstrahlung source only dipole and to a much lesser
extent also electric quadrupole excitations will be induced.
Secondly, the theoretical model should be able to distinguish between
the degrees of freedom of the collective phonons and the additional
particle degrees of freedom which open extra possible excitation
channels and which have no counterparts in the even-even nuclei. 

Electric dipole excitations to a particle two-phonon multiplet were
for the first time identified in $^{143}$Nd \cite{Zilg93}. In the
energy region between 2.8 and 3.8~MeV, 13 levels were observed for
which an underlying particle two-phonon $f_{7/2} \otimes [2^{+}_{1}
\otimes 3^{-}_{1}]$ structure was suggested.  The observed
fragmentation and $B(E$1)$\uparrow$ strength distribution could be
reproduced in a phenomenological simple core coupling model based on
quadrupole-quadrupole coupling \cite{Herz95a}.  Moreover, the summed 
experimental $B(E$1)$\uparrow$ strength between 2.8 and 3.8 MeV agrees
within the statistical error with the known $B(E$1)$\uparrow$ strength
of the two-phonon $[2^{+}_{1} \otimes 3^{-}_{1}]_{1^{-}}$ state in the 
neighbouring even-mass $^{142}$Nd nucleus.  It was concluded that the
unpaired neutron in its $f_{7/2}$ orbital, outside the closed major N
= 82 shell, couples extremely weakly to the two-phonon $[2^{+}_{1} \otimes
3^{-}_{1}]$ quintuplet and plays the role of a pure spectator.
Later on, similar NRF-experiments performed on $^{139}$La and $^{141}$Pr
\cite{Herz95c} revealed also a large fragmentation of the electric
dipole strength, but in both cases less than 40\% of the two-phonon
$B(E$1)$\uparrow$ strength in the neighbouring even-mass $^{138}$Ba,
$^{140}$Ce and $^{142}$Nd nuclei was observed.  In $^{139}$La and
$^{141}$Pr the odd proton in the partly filled shell couples more strongly
to the two-phonon quintuplet.  Intermediate cases have been observed
in the open shell nuclei $^{113}$Cd \cite{Geig94} and $^{133}$Cs
\cite{Bess97}.

The odd-mass $^{117}$Sn nucleus was chosen to investigate the
fragmentation of the well-known two-phonon $B(E$1)$\uparrow$ strength
from its even-mass neighbours $^{116}$Sn and $^{118}$Sn.  The NRF
technique was used for obvious reasons. In $^{117}$Sn, the unpaired
neutron is situated halfway the major N = 50 and 82 shells.  As an 
interesting property, the ground state spin $J^{\pi}_{0} = 1/2^{+}$ limits 
the possible dipole excitations to levels with a spin $J = 1/2$ or
$3/2$ and electric quadrupole excitations can only occur to states
with a spin and parity $J^{\pi} = 3/2^{+}$ and $5/2^{+}$. For the
first time, calculations within the QPM are carried out in a complete
configuration space for an odd-mass spherical nucleus.  Our first
results on the experimentally observed fragmentation and on the
performed calculations were described in a previous letter paper
\cite{Pono99}.  In the present paper a comprehensive discussion of the
experimental and theoretical aspects of our work will be presented.

\section{Experimental method and setup}
The nuclear resonance fluorescence technique or the resonant
scattering of real photons off nuclei as described in many reviewing
articles e.g.\ \cite{Knei96,Metz59,Skor75}, was applied to investigate the
$^{117}$Sn nucleus.  The main advantages of this real photon probe
consist of the highly selective excitation of levels, as pointed out
in the introduction, and the possibility of a completely model
independent analysis of the data. The use of HP Ge detectors to
detect the resonantly scattered photons, allows the observation of
individual levels and the determination of their excitation energies $E_{x}$
with a precision better than 1 keV.  The total elastic scattering
cross section $I_{S}$, energy integrated over a single resonance and
integrated over the full solid angle equals \cite{Knei96}:
\begin{equation}
I_{S} = g \left(\frac{\pi \hbar c}{E_{x}}\right)^2 
\frac{\Gamma^{2}_{0}}{\Gamma} 
\label{eqis}
\end{equation}
with $\Gamma_{0}$ and $\Gamma$ the ground state and total transition
width and $g$ a statistical factor depending on the ground
state spin $J_{0}$ and the spin $J$ of the excited level: 
\begin{equation}
g = \frac{2J + 1}{2J_{0} + 1} \ \ .
\end{equation}
In our experiments, the scattering cross sections of the observed
levels in $^{117}$Sn are determined relative to the $^{27}$Al calibration
standard.  The spectral shape of the incoming bremsstrahlung flux is
fitted using the Schiff formula for thin targets and the cross
sections for the transitions in $^{117}$Sn are extracted relatively to
the cross sections of well-known transitions in $^{27}$Al
\cite{Piet95}.  For even-even nuclei, the spin $J$ of the excited
level is easily obtained from the measured angular 
distribution of the resonantly scattered photons.  A clear distinction
between dipole and quadrupole transitions can be made by comparing the
observed $\gamma$ intensities at the scattering angles of $90^{\circ}$
and $127^{\circ}$ \cite{Knei96}.  However, for odd-mass nuclei the
extraction of limited spin information is only possible for nuclei as
$^{117}$Sn with a ground state spin $J_{0}$ of 1/2
\cite{Geig94}. Higher half integer ground state spins lead to nearly 
isotropic angular distributions.  From an evaluation of the 
angular distribution function $W$ for the involved spin sequence,
scattering angle and mixing ratio $\delta$, the following results are
obtained: \newline   
$\frac{1}{2} \rightarrow \frac{1}{2} \rightarrow \frac{1}{2}$
\begin{equation} 
W(90^{\circ}) = W(127^{\circ}) = W(150^{\circ}) = 1 
\end{equation}
$\frac{1}{2} \rightarrow \frac{3}{2} \rightarrow \frac{1}{2}$
\begin{equation} 
\frac{W(90^{\circ})}{W(127^{\circ})} = 0.866; \ \frac{W(90^{\circ})}
{W(150^{\circ})} = 0.757; \ \ \delta = 0, \pm \infty 
\end{equation}
$\frac{1}{2} \rightarrow \frac{5}{2} \rightarrow \frac{1}{2}$
\begin{equation}
\frac{W(90^{\circ})}{W(127^{\circ})} = 1.168; \ \frac{W(90^{\circ})}
{W(150^{\circ})} = 0.842; \ \ \delta = 0 \ \ .
\end{equation}
The scattering angles correspond to those used in our
experiments. The angular distribution functions $W$ are independent of
the parity of the excited level.  In Fig.~\ref{sn117angdis} the
expected angular correlation function ratios
$W(90^{\circ})/W(150^{\circ})$ are plotted versus the ratio
$W(90^{\circ})/W(127^{\circ})$ for different values of the mixing
ratio $\delta$ and for the three possible induced spin sequences.  A
level with spin 1/2 can only be excited from the ground state via a
dipole transition and hence no mixing of multipolarities is possible
in a 1/2 -- 1/2 -- 1/2 spin cascade.  Such a cascade has an isotropic
angular distribution function $W$.  The square in
Fig.~\ref{sn117angdis} represents the unique location in this figure 
where 1/2 -- 1/2 -- 1/2 spin sequences can be found.  In the case of a 
1/2 -- 3/2 -- 1/2 spin sequence, assuming a positive parity for the
1/2 state (as is the case in $^{117}$Sn), a mixing between $E1$ and $M2$
or $M1$ and $E2$ transitions is possible depending on the parity of the
excited 3/2 level.  However, $M2$ transitions can not be observed in
NRF experiments.  For a pure $E1$ ($M1$) transition, corresponding to  
a mixing ratio $\delta$ equal 0, or a pure $M2$ ($E2$) transition,
with mixing ratio $\delta$ equal $\pm \infty$,  to the 3/2 level with
a negative (positive) parity, the same two values for the
ratios $W(90^{\circ})/W(127^{\circ})$ and
$W(90^{\circ})/W(150^{\circ})$ are obtained. This is represented by
the triangle in 
Fig.~\ref{sn117angdis}.  For mixing ratios varying between  
0 and $\pm \infty$ the point representing the couple of ratios 
$W(90^{\circ})/W(127^{\circ})$ and $W(90^{\circ})/W(150^{\circ})$ moves
along the solid line in Fig.~\ref{sn117angdis}. For a 1/2 -- 5/2 -- 1/2 spin
sequence, transitions with multipolarities L = 2 and L = 3 are
theoretically possible.  However, $M2$, $E3$ and $M3$ excitations can
be excluded to be observed in our NRF-experiments because they have
scattering cross sections far below the sensitivity of our
setup. Therefore, only a pure $E2$ excitation to a 5/2 level 
can be observed.  The star in Fig.~\ref{sn117angdis} shows where
these 1/2 -- 5/2 -- 1/2 spin sequences can be expected. For the
strongest transitions observed with a high statistical precision, the
experimental uncertainties on the ratios
$W(90^{\circ})/W(127^{\circ})$ and $W(90^{\circ})/(150^{\circ})$ will
be small enough to suggest the induced half integer spin sequence. In
most cases the statistical accuracy will not allow to determine the
spin sequence and hence the statistical spin factor $g$ can not be
determined.  

Compton polarimetry \cite{Schl94} and the scattering of linearly
polarized off-axis bremsstrahlung \cite{Berg87,Gova94a} represent two
useful techniques to determine the parities of the excited levels in
even-even nuclei \cite{Knei96}.  For odd-mass nuclei, the measured
azimuthal asymmetry of the resonantly scattered photons in both
methods will nearly vanish because of the half integer spin
sequences and strongly depend on the mixing ratio $\delta$. The lower
statistics in these experiments do not allow to distinguish between
the different possible cases.   For a further analysis of the data, it
will be assumed that all observed levels are populated via pure
electric dipole excitations and that they do not have any decay
branchings (unless observed otherwise) to intermediate lower-lying
levels ($\Gamma_{0}/\Gamma  = 1$). In this case, the product of the
ground state transition width and the spin factor g can immediately be 
extracted from the measured scattering cross section $I_{S}$ (Eq.\
\ref{eqis}) and the reduced electric dipole excitation probability
is given by:     
\begin{equation}
B(E1)\uparrow = \frac{2.866}{3} \cdot \frac{g \cdot \Gamma_{0}}
{E^{3}_{x}} \ \ (10^{-3} \cdot e^{2}fm^{2})
\end{equation}
with the ground state transition width $\Gamma_{0}$ in meV and
the excitation energy $E_{x}$ in MeV.  The deduced
$B(E$1)$\uparrow$ strength in our NRF-experiments, includes
automatically the statistical factor g and hence can immediately be
compared with the observed $B(E$1)$\uparrow$ strength in other nuclei
or calculated from theoretical models. 

The experiments were performed at the NRF-facility of the 4.3~MV
Dynamitron accelerator of the Stuttgart University.  The energy of the 
electron beam was 4.1~MeV and the beam current was limited to about
250~$\mu$A due to the thermal characteristics of the bremsstrahlung
production target and to avoid too high count rates in the
detectors.  A setup consisting of 3 HP Ge detectors, installed at 
scattering angles of 90$^{\circ}$, 127$^{\circ}$ and 150$^{\circ}$
each with an efficiency $\epsilon$ of 100\% (relative to a 3" x 3"
NaI(Tl) detector), was used to measure the total elastic scattering cross
sections of the levels in $^{117}$Sn.  These HP Ge detectors allow to
detect the resonantly scattered photons with a high sensitivity and a
very good energy resolution. Two metallic Sn disks with a diameter of
1~cm, a total weight of 1.649~g, and an isotopic enrichment of 92.10\%
in $^{117}$Sn were irradiated during five days.  Two $^{27}$Al disks
with a total amount of 0.780~g were alternated with the two $^{117}$Sn
disks for calibration purposes.  

\section{Results}
Part of the recorded $^{117}$Sn ($\gamma$,$\gamma'$) spectrum
(2.6 $\leq$ $E_{\gamma}$ $\leq$ 3.6~MeV) is shown in
Fig.~\ref{sn117spec} together with the spectra of its even-even
$^{116}$Sn and $^{118}$Sn neighbours. In our previous studies on
$^{116}$Sn and $^{118}$Sn \cite{Brys99}, we found that the
$E1$ strength in this energy region below 4~MeV is mostly concentrated
in the two-phonon $[2^{+}_{1} \otimes 3^{-}_{1}]_{1^{-}}$ state.  In
the spectra of the even-even nuclei, the dominating peak at an energy
of about 3.3~MeV corresponds to the deexcitation of this two-phonon
1$^{-}$ state into the ground state.  In comparison to the spectra of
the even-even Sn nuclei, the $^{117}$Sn spectrum contains a lot of
$\gamma$ transitions superimposed on the smooth background in the
vicinity of the two-phonon 1$^{-}$ states .  The peaks
stemming from the deexcitation of the two-phonon $[2^{+}_{1} \otimes
3^{-}_{1}]_{1^{-}}$ states in $^{116,118,120}$Sn have also been
observed in the spectrum due to the small admixtures of
0.86\%, 5.81\% and 0.76\% in the target. In Fig.\
\ref{sn117spec}b only the peak for $^{118}$Sn can be
clearly observed due to the scale used. All observed
$\gamma$ transitions in $^{117}$Sn are listed in Table
\ref{sn117results} with the corresponding level excitation energies, the
measured total elastic scattering cross sections $I_{S}$, the
extracted transition width ratios $g \cdot
\frac{\Gamma^{2}_{0}}{\Gamma}$ (depending on the statistical
spin-factor g) and deduced electric excitation probabilities
$B(E$1)$\uparrow$.  The total elastic scattering cross sections has
been determined from a summed spectrum over the three scattering
angles. The three HP Ge-detectors have nearly the same efficiency and
the summed angular distributions over the three detectors equals three 
for all possible spin cascades.  This method allows us to observe also
some weaker lines.

In some cases, the observed angular distribution ratios provide an
indication of the spin $J$ of the photo-excited levels. These results
are summarized in Table \ref{sn117spin}.  In the first column the 
energies of these levels are given.  In the next columns the observed
angular distribution ratios $W(90^{\circ})/W(127^{\circ})$ and
$W(90^{\circ})/W(150^{\circ})$ and the suggested spin are
presented. The experimentally observed angular distribution ratios
$W(90^{\circ})/W(150^{\circ})$ versus $W(90^{\circ})/(127^{\circ})$
are included in Fig.\ \ref{sn117angdis} (points with error
bars).  Two groups of $\gamma$ transitions can clearly be observed.
For a first group of 5 levels, located around the triangle, 
a probable spin assignment of 3/2 can be given.  A second group of
6 levels can be found around the square.  Here an assignment of 1/2
(3/2) can be given.  We prefer $J = 1/2$ for these states because
we suppose that strong transitions should have an $E1$ character.  These
$E1$ transitions have an isotropic distribution only for excited
states with $J = 1/2$.  However we can not exclude $J = 3/2$ (negative 
or positive parity) as transitions with a mixing ratio $\delta$ around
-3.73 or 0.27 also lead to an isotropic distribution.  All the spins
given in Table \ref{sn117spin} were assigned at least at a statistical
significance level of $1 \sigma$.  Under these conditions, no levels
with a spin of 5/2 were found.  

Five of the observed $\gamma$ transitions are
probably due to inelastic deexcitations of a level to the well-known
first $3/2^{+}$ state in $^{117}$Sn at 158.562(12)~keV \cite{Blac87}.
They are summarized in Table \ref{sn117inelas}.  In this Table the
level excitation energies $E_{x}$, the energies $E_{\gamma}$ of the
deexciting $\gamma$ transitions,  the relative $\gamma$ intensities
$I_{\gamma}$, the ground state transition widths $g \cdot \Gamma_{0}$
and reduced electric dipole excitation probabilities corrected for the
observed inelastic decays, are given. All levels are included for
which holds: 
\begin{equation}
E_{i} - (E_{x} - E_{158}) < \Delta\!E 
\end{equation}
with:
\begin{equation}
\Delta\!E = \sqrt{(\Delta\!E_{x})^{2} + (\Delta\!E_{i})^{2} + 
(\Delta\!E_{158})^{2}}
\end{equation}
and $E_{x}$, $E_{i}$, $E_{158}$ and $\Delta\!E_{x}$, $\Delta\!E_{i}$,
$\Delta\!E_{158}$ the excitation energies of the
level, the energy of the inelastic $\gamma$ transition and the
first $3/2^{+}$ state and their respective uncertainties. Using the
above mentioned rule for detecting inelastic decays
two other candidates were found.  The $\gamma$ ray with the energy of
3560.5 keV can be due to an inelastic transition of one line from the
multiplet at 3719.8 keV to the first $3/2^{+}$ state.  Also the
$\gamma$ ray with the energy of 2986.7 keV fits into the energy relation
with the observed level at 3144.9 keV.  However, when this
$\gamma$ ray is seen as completely inelastic, an unreasonable low
branching ratio $\Gamma_{0}/\Gamma$ of 12\% for the 3144.9 keV level
is obtained.  Both cases are not considered in
Table~\ref{sn117inelas}.  Certainty about the observed probable
inelastic $\gamma$ transitions requires time coincidence
measurements.  With the above mentioned method, our experimental
results show no further candidates for inelastic transitions to other
intermediate observed levels.

\section{Discussion}

According to the phenomenological core coupling model, the level scheme
for $^{117}$Sn can be obtained from the coupling of the odd 3s$_{1/2}$
neutron to the $^{116}$Sn core.  This is
schematically shown in Fig.~\ref{sn117scheme}.  Each level in
$^{116}$Sn (except for $J = 0$ levels) gives rise to two new levels due
to the spin 1/2 of the odd 
neutron.  The low-lying level scheme of $^{116}$Sn is dominated by the
strong quadrupole ($2^{+}$) and octupole ($3^{-}$) vibrational states, 
typical for a spherical semi-magic nucleus.  The coupling of the
3s$_{1/2}$ neutron to the first $2^{+}$ state in $^{116}$Sn results in 
two new levels $[3s_{1/2} \otimes 2^{+}_{1}]_{3/2^{+}}$ and
$[3s_{1/2} \otimes 2^{+}_{1}]_{5/2^{+}}$ which can be excited in NRF
via $M1$ and $E2$ transitions (solid lines in Fig.~\ref{sn117scheme}).
The similar doublet consisting of the $[3s_{1/2} \otimes
3^{-}_{1}]_{5/2^{-}}$ and $[3s_{1/2} \otimes 3^{-}_{1}]_{7/2^{-}}$
states requires $M2$ and $E3$ excitations which are not observable
in NRF (dashed lines).  The main aim of our NRF-experiments was to
search for levels belonging to the $3s_{1/2} \otimes [2^{+}_{1}
\otimes 3^{-}_{1}]$ multiplet which can be populated via electric
dipole transitions.  When the quadrupole-octupole coupled two-phonon
quintuplet is built on top of the 1/2$^{+}$ ground state, a multiplet
of 10 negative parity states is obtained of which 3 levels can be
excited via $E1$ transitions (solid lines in Fig.~\ref{sn117scheme}).
In this simple model, these three transitions carry the complete
$B(E$1)$\uparrow$ strength. In Fig.~\ref{sn117is}b
the obtained total scattering cross sections $I_{S}$ for the
photo-excited levels in $^{117}$Sn (with exclusion of the lines which
are probable due to inelastic scattering given in  Table \ref{sn117inelas})
and in its even-even neighbours $^{116}$Sn and $^{118}$Sn are
plotted. The total scattering 
cross section for the excitation of the two-phonon $[2^{+}_{1} \otimes
3^{-}_{1}]$ states in $^{116}$Sn and $^{118}$Sn has been reduced by a
factor of 3.  A strong fragmentation of the strength has been observed
in $^{117}$Sn compared to its even-even neighbours.  It is already
clear from the observed fragmentation of the strength that a
phenomenological core coupling model, which was successful in
describing the observed strength in $^{143}$Nd, will be insufficient
in our case. Due to a lack of spin and parity information of the
photo-excited levels in $^{117}$Sn in our NRF-experiment and due to a
lack of experimental data from other investigations \cite{Blac87}, we
need to turn to a more elaborated theoretical interpretation to get
more insight.   

\subsection{QPM formalism for odd-mass nuclei}

The Quasiparticle phonon model (QPM) was already successful in
describing collective properties in even-even mass nuclei
\cite{Solo92}. Recently, the QPM has been applied to describe the
position and the $E1$ excitation probability of the lowest $1^-$ state
in the even-even $^{116-124}$Sn isotopes \cite{Brys99}. This state
has a two-phonon character with a contribution of the $[2^+_1 \otimes
3^-_1]_{1^-}$ configuration of 96-99\%. For odd-mass nuclei, this
model was used to describe the fragmentation of deep-lying hole and
high-lying particle states \cite{Vdov85,Gale88} and the
photo-production of isomers \cite{Pono90,Hube93,Carr93}. It has
already been applied to calculate the absolute amount of strength in
$^{115}$In \cite{Cose95}, but up till now it has not been extended to
describe and understand the high fragmentation of the strength and the
distribution of the $B(E$1)$\uparrow$, $B(M$1)$\uparrow$, and
$B(E$2)$\uparrow$ strength in the energy region below 4~MeV. 

General ideas about the QPM and its formalism to describe the excited
states in odd-mass spherical nuclei with a wave function which
includes up to ``quasiparticle $\otimes$ two-phonon'' configurations are
presented in review articles \cite{Vdov85,Gale88}.  It is extended here
by including ``quasiparticle $\otimes$ three-phonon'' configurations as well.
A Woods-Saxon potential is used in the QPM as an average field for
protons and neutrons. Phonons of different multipolarities and
parities are obtained by solving the RPA equations with a separable form
of the residual interaction including a Bohr-Mottelson form factor. 
The single-particle spectrum and phonon basis are fixed from calculations 
in the neighboring even-even nuclear core, i.e. in $^{116}$Sn
\cite{Brys99} when $^{117}$Sn nucleus is considered.

In our present calculations the wave functions of the ground state and 
the excited states are mixtures of different ``quasiparticle $\otimes$
$N$-phonon'' ($[qp \otimes N ph]$) configurations, where $N$ =0, 1, 2,
3: 
\begin{eqnarray}
&&\Psi^{\nu}(JM) =
\left \{ C^{\nu}(J) \alpha^+_{JM}~+~
\sum_{j \beta_1} S^{\nu}_{j \beta_1}(J) 
[\alpha^+_j Q^{+}_{\beta_1}]_{JM} \right.
\nonumber \\
&+&
\sum_{j \beta_1 \beta_2} \frac{D^{\nu}_{j \beta_1 \beta_2}(J)
[\alpha^+_j Q^{+}_{\beta_1}  Q^{+}_{\beta_2}]_{JM}}
{\sqrt{1+\delta_{\beta_1 \beta_2}}}
\label{wf} 
   \\
&+& \left.
\sum_{j \beta_1 \beta_2 \beta_3} \frac{T^{\nu}_{j \beta_1 \beta_2
\beta_3}(J)
[\alpha^+_j Q^{+}_{\beta_1} Q^{+}_{\beta_2} Q^{+}_{\beta_3}]_{JM} }
{\sqrt{1+\delta_{\beta_1 \beta_2} + \delta_{\beta_1 \beta_3}
+ \delta_{\beta_2 \beta_3}
+ 2 \delta_{\beta_1 \beta_2 \beta_3}}}
 \right \}
\mid \rangle _{g.s.}
\nonumber
\end{eqnarray}
where the coefficients $C$, $S$, $D$ and $T$ describe a contribution
of each configuration to a norm of the wave function. We use the
following notations $\alpha^+$ and $Q^+$ for the coupling between the
creation operators of quasiparticles and phonons.  
\[
[\alpha^+_j Q^{+}_{\lambda i}]_{JM}  = 
 \sum_{m \mu} C_{j m \lambda \mu}^{JM} 
\alpha^+_{jm} Q^{+}_{\lambda \mu i}\ ,
\]
\[
[\alpha^+_j Q^{+}_{\beta_1} Q^{+}_{\beta_2} Q^{+}_{\beta_3}]_{JM}
= 
\sum_{\lambda_1 \lambda_2}
[\alpha^+_j [Q^{+}_{\beta_1} [Q^{+}_{\beta_2} 
Q^{+}_{\beta_3}]_{\lambda_1}]_{\lambda_2}]_{JM}~,
\]
\begin{equation}
[Q^{+}_{\lambda_1 i_1} Q^{+}_{\lambda_2 i_2}]_{\lambda \mu}
= 
\sum_{\mu_1 \mu_2} C_{\lambda_1 \mu_1 \lambda_2 \mu_2}^{\lambda \mu} 
 Q^{+}_{\lambda_1 \mu_1 i_1}  Q^{+}_{\lambda_2 \mu_2 i_2} 
\end{equation}
where $C$ are Clebsh-Gordon coefficients.  Quasiparticles are
characterized by their shell quantum numbers $jm \equiv |nljm>$ with a
semi-integer value of the total angular momenta $j$. They are the result
of a Bogoliubov transformation from particle creation (annihilation)
$a^+_{jm}$ ($a_{jm}$) operators: 
\begin{equation}
a_{jm}^{+}\,=\,u_{j}\alpha_{jm}^{+}\,+\,(-1)^{j-m}v_{j}\alpha_{j-m}~.
\label{qua}
\end{equation}
The quasiparticle energy spectrum and the occupation number coefficients
$u_{j}$ and $v_{j}$ in Eq.~(\ref{qua}) are obtained in the QPM by
solving the BCS equations separately for neutrons and protons.

Phonons with quantum numbers $\beta \equiv |\lambda \mu i>$ are linear
superpositions of two-quasiparticle configurations:
\begin{eqnarray}
Q^{+}_{\lambda \mu i}\,&=&\,\frac{1}{2}
\sum_{\tau}^{n,p}\sum_{jj'} \left \{ \psi_{jj'}^{\lambda i}
[\alpha^{+}_{j} \alpha^{+}_{j'}]_{\lambda \mu}
\right .
\nonumber \\
&-&\left. (-1)^{\lambda - \mu} \varphi_{jj'}^{\lambda i}
[\alpha_{j'}\alpha_{j}]_{\lambda -\mu} \right \}~.
\label{ph}
\end{eqnarray}
A spectrum of phonon excitations is obtained by solving the RPA
equations for each multipolarity $\lambda$ which is an integer value.
The RPA equations also yield forward (backward) $\psi_{jj'}^{\lambda i}$
($\varphi_{jj'}^{\lambda i}$) amplitudes in definition (\ref{ph}): 
\begin{equation}
{\psi \choose \varphi}^{\lambda i}_{j j'} (\tau) =
\frac{1} {\sqrt{ \cal{Y} _{\tau} ^{\lambda \mit{i}} }}
\cdot
\frac{f^{\lambda}_{j j'}(\tau) (u_j v_{j'}+ u_{j'} v_j)}
{\varepsilon_{j} + \varepsilon_{j'} \mp \omega_{\lambda i}}
\end{equation}
where $\varepsilon_{j}$ is a quasiparticle energy, $\omega_{\lambda i}$
is the energy needed for the excitation of an one-phonon
configuration, $f_{jj'}^{\lambda}$ is a reduced single-particle matrix
element of residual forces, and the value $\cal{Y} _{\tau} ^{\lambda
\mit{i}}$ is determined from a normalization condition for the phonon
operators:
\begin{equation}
\langle | Q_{\lambda \mu i} Q_{\lambda \mu i}^+ | \rangle_{ph} =
 \sum_{\tau}^{n,p} \sum_{jj'} \left \{ ( \psi^{\lambda i}_{jj'})^{2} -
(\varphi^{\lambda i}_{jj'} \right )^{2} \}  = 1~.
\end{equation}
The phonon's index $i$ is used to distinguish between phonon excitations 
with the same multipolarity but with a difference in energy and structure.
The RPA equations yield both, collective- (e.g. $2^+_1$ and $3^-_1$), and 
weakly-collective phonons. The latter correspond to phonons for which
some specific two-quasiparticle configuration is dominant in
Eq.~(\ref{ph}) while for other configurations $\psi_{jj'}^{\lambda i},
\varphi_{jj'}^{\lambda i} \approx 0$.

When the second, third, etc. terms in the wave function of
Eq.~(\ref{wf}) are considered, phonon excitations of the core
couple to a quasiparticle at any level of the average field, 
not only at the ones with the quantum
numbers $J^{\pi}$ as for a pure quasiparticle configuration.
 It is only necessary that all
configurations in Eq.~(\ref{wf}) have the same total spin and
parity. To achieve a correct position of the $[qp \otimes 2 ph]$
configurations, in which we are especially interested in these
studies, $[qp \otimes 3 ph]$ configurations are important. The
excitation energies and the contribution of the different components from
the configuration space to the structure of each excited state  
(i.e. coefficients $C$, $S$, $D$ and $T$ in Eq.~(\ref{wf})) are
obtained by a diagonalization of the model Hamiltonian on a set of
employed wave functions. The coupling matrix elements between the
different configurations in the wave functions of Eq.~(\ref{wf}) in
odd-mass nuclei are calculated on a microscopic footing, making use of
the internal fermion structure of the phonons and the model
Hamiltonian. For example, the interaction matrix element between the 
$[qp \otimes  1 ph]$ and the $[qp \otimes  2 ph]$ configurations has
the form (see, Ref.~\cite{Gale88}):
\begin{eqnarray}
&&<[\alpha_{jm} Q_{\lambda \mu i}]_{JM} | H |
[\alpha^+_{j'm'} [Q^+_{\lambda_1 \mu_1 i_1} 
Q^+_{\lambda_2 \mu_2 i_2}]_{I M'} ]_{JM}>
\nonumber \\[2mm]
&&=  \delta_{jj'} \delta_{\lambda I} 
U^{\lambda_2 i_2}_{\lambda_1 i_1} (\lambda i)
- (-)^{j'+\lambda+I} 2 \sqrt{(2j+1) (2I+1)}
\nonumber \\
&&\times \left[ (-)^{\lambda_1} \delta_{\lambda \lambda_1} 
\left\{ \begin{array}{ccc}
\lambda_{2} & \lambda_{1} & I \\
J  & j  & j' \end{array} \right\}
\Gamma(j j' \lambda_2 i_2) \right .
\nonumber \\
&& + \left . 
(-)^{\lambda_2} \delta_{\lambda \lambda_2} 
\left\{ \begin{array}{ccc}
\lambda_{1} & \lambda_{2} & I \\
J  & j  & j' \end{array} \right\}
\Gamma(j j' \lambda_1 i_1)
\right ]
\label{int1}
\end{eqnarray}
where $H$ is a model Hamiltonian, $U^{\lambda_2 i_2}_{\lambda_1 i_1}
(\lambda i)$ is an interaction matrix element between one- and
two-phonon configurations in the neighbouring even-mass nucleus ($U$
is a complex function of phonon's amplitudes $\psi$ and $\varphi$ and $f_{j
j'}^{\lambda}$; its explicit form can be found
in Ref.~\cite{Bert99}) and $\Gamma$ is an interaction matrix element
between quasiparticle $\alpha^+_{JM}$ and quasiparticle-phonon
$[\alpha^+_{jm} Q^+_{\lambda \mu i}]_{JM}$ configurations, it is equal
to: 
\begin{equation}
\Gamma(J j \lambda i)
= \sqrt{\frac{2\lambda +1}{2J+1}} \, \,
\frac{f_{J j}^{\lambda} (u_J u_j - v_j v_J)}
{\sqrt{{\cal Y}^{\lambda i}_{\tau}}}
~.
\label{int2}
\end{equation}
 
Equations (\ref{int1},\ref{int2}) are obtained by applying
the exact commutation relations between the phonon and quasiparticle
operators: 
\begin{eqnarray}
[\alpha_{jm}, Q^+_{\lambda \mu i} ]_{\_} &=& \sum_{j'm'}
\psi^{\lambda i}_{jj'} C^{\lambda \mu}_{jm j'm'} \alpha^+_{j'm'}~,
\label{a-q-c}
\\
\nonumber 
[\alpha^+_{jm}, Q^+_{\lambda \mu i} ]_{\_} &=&
(-1)^{\lambda - \mu} \sum_{j'm'}
\varphi^{\lambda i}_{jj'} C^{\lambda -\mu}_{jm j'm'} \alpha_{j'm'} 
\ \ .
\end{eqnarray}
The exact commutation relations between the phonon operators 
$Q_{\lambda \mu i}$ and $Q^{+}_{\lambda' \mu' i'}$
\begin{eqnarray}
&&[Q_{\lambda \mu i}, Q^{+}_{\lambda' \mu' i'} ]_{\_}
= 
\delta_{\lambda \lambda'} \delta_{\mu \mu'}
\delta_{i i'} 
\nonumber
\\
&-& \sum_{\scriptstyle jj'j_{2} \atop \scriptstyle m m'm_{2}}
\alpha^{+}_{jm} \alpha_{j'm'}
\times
\left \{
\psi^{\lambda i}_{j'j_{2}} \psi^{\lambda'i'}_{jj_{2}}
C_{j'm' j_2m_2}^{\lambda \mu}
C_{j m  j_2m_2}^{\lambda' \mu'}  \right.  \nonumber \\
&-& \left. (-)^{\lambda + \lambda'+\mu + \mu'}
\varphi^{\lambda i}_{jj_{2}} \varphi^{\lambda'i'}_{j'j_{2}}
C_{j m  j_2m_2}^{\lambda -\mu}
C_{j'm' j_2m_2}^{\lambda' -\mu'}
\right \}
\label{q-q-c}
\end{eqnarray}
are used to calculate the interaction matrix elements $U$ in even-even  
nuclei. 

The interaction matrix elements between the $[qp \otimes  2 ph]$ and 
the $[qp \otimes  3 ph]$ configurations have a structure similar to
(\ref{int1}). We do not provide them here because of their complexity.  
But even Eq.~(\ref{int1}) shows that an unpaired quasiparticle does 
not behave as a spectator but modifies the interaction between the complex  
configurations compared to an even-mass nucleus (see second term in
this equation). This takes place because the phonons possess an 
internal fermion structure and the matrix elements $\Gamma$ correspond to
an interaction between an unpaired quasiparticle and the two-quasiparticle
configurations composing the phonon operator.

It should be pointed out that in the present approach interaction
matrix elements are calculated in first order perturbation
theory. This means that any $[qp \otimes N ph]$ configuration
interacts with the $[qp \otimes (N \pm 1) ph]$) ones, but its coupling
to $[qp \otimes (N \pm 2) ph]$ configurations is not included in this
theoretical treatment. The omitted couplings have non-vanishing
interaction  matrix elements only in second order perturbation
theory. They are much smaller than the ones taken into account and
they are excluded from our consideration for technical reasons. An
interaction with other $[qp \otimes N ph]$ configurations is taken
into account while treating the Pauli principle corrections. In a
calculation of the self-energy of the complex configurations we employ
a model Hamiltonian written in terms of quasiparticle operators and
exact commutation relations (\ref{a-q-c},\ref{q-q-c}) between
quasiparticle and phonon operators. In this case, we obtain a ``Pauli
shift correction'' for the energy of a complex configuration from the
sum of the energies of its constituents. Also, when considering
complex configurations their internal fermion structure is analyzed
and the ones which violate the Pauli principle are excluded from the 
configuration space.  
Pauli principle corrections have been 
treated in a diagonal approximation (see, Ref.~\cite{Vdov85} for details).

In the actual calculations, the phonon basis includes the phonons with
multipolarity and parity $\lambda^{\pi} =~1^{\pm},~2^+,~3^-$ and
$4^+$. Several low-energy phonons of each multipolarity are included
in the model space. The most important ones are the first collective
$2^+,~3^-$ and $4^+$ phonons and the ones which form the giant dipole
resonance (GDR). Non-collective low-lying phonons of an unnatural
parity and natural parity phonons of higher multipolarities are of a
marginal importance. To make realistic calculations possible one has
to truncate the configuration space. We have done this on the basis of
excitation energy arguments. All $[qp \otimes 1 ph]$ and $[qp \otimes
2 ph]$ with $E_x \le 6$~MeV, and $[qp \otimes 3 ph]$ with $E_x \le
8$~MeV configurations are included in the model space. The only
exceptions are $[J_{g.s.} \otimes 1^-]$ configurations which have not
been truncated at all to treat a core polarization effect due to the
coupling of low-energy dipole transitions to the GDR on a microscopic
level. Thus, for electric dipole transitions we have no renormalized
effective charges and used $e^{{\rm eff}}(p)=(N/A)\,e$ and $e^{{\rm
eff}}(n)=-(Z/A)\,e$ values to separate the center of mass motion. For
$M1$ transitions we use $g_s^{{\rm eff}}= 0.64g_s^{{ \rm free}}$ as
recommended in Ref.~\cite{Cose99}. By doing this all the important
configurations for the description of low-lying states up to 4~MeV are
included in the model space. The dimension of this space depends on
the total spin of the excited states, and it varies between 500 and
700 configurations. 

\subsection{Comparison between experimental data and QPM calculations}

Since only $E1$, $M1$ and $E2$ transitions can be observed in the
present experiment, the discussion of the properties of the excited
states will be restricted to states with $J^{\pi} =
1/2^{\pm},~3/2^{\pm}$ and $5/2^+$. As the parities of the decaying
levels are unknown and the spin could be assigned for only a few
levels, the best quantity for the comparison between the theoretical
predictions and the experimental results are the total integrated cross
sections $I_s$. The theoretical reduced excitation probabilities
$B(\pi L$)$\uparrow$ can be transformed into $I_{S}$ values via the
following relation:
\begin{equation} 
I_{S} = \frac{8 \pi^{3} (L + 1)}{L[(2L+1)!!]^{2}} 
\left( \frac{E_{x}}{\hbar c} \right)^{2L-1} \cdot B(\pi L)\uparrow 
\cdot \frac{\Gamma_{0}}{\Gamma_{tot}}~,
\label{eq}
\end{equation} 
where $E_x$ is the excitation energy, $L$ the multipolarity of the
transition and $\Gamma_{0}$ denotes the partial ground state decay
width.  The obtained $I_{S}$ values for the elastic transitions are
plotted in Fig.~\ref{sn117is}c and compared with the results of our
$(\gamma,\gamma')$ experiments given in Fig.~\ref{sn117is}b. The
inelastic decays are accounted for in the total decay widths
$\Gamma_{tot}$. Details concerning the calculations and branching
ratios will be discussed below. Supporting the experimental findings
our calculations also produce a strong fragmentation of the
electromagnetic strength.  The strongest transitions have $E1$
character, but also $E2$ and $M1$ excitations yield comparable cross
sections. The total cross section $I_{S}$ is disentangled into its
$E1$, $M1$ and $E2$ components in Fig.~\ref{fig2}b,c,d, and compared
to the calculated $I_s$ values of the core nucleus, $^{116}$Sn
(Fig.~\ref{fig2}a). The calculated sum of the total cross sections of the
plotted $E1$, $M1$ and $E2$ transitions in Fig~\ref{fig2}b-d equals 73,
37 and 42 eVb. The summed experimental elastic cross sections,
shown in Fig.~\ref{sn117is}b, equals 133 (22) eVb and agrees within 15\%
with the theoretically predicted value of 152 eVb. 

Although the experimentally observed levels do not match in detail
with the calculated level scheme one by one, some interesting general
conclusions can be drawn. The most essential differences in the
electromagnetic strength distribution over low-lying states in
even-even $^{116,118}$Sn and odd-mass $^{117}$Sn take place for the
electric dipole transitions.  The reason becomes clear by considering
which states can be excited from the ground state by $E1$
transitions. In the even-even core there is only one $1^-$
configuration with an excitation energy below 4~MeV (thick line with
triangle in Fig.~\ref{fig2}a).  It has a $[2^+_1 \otimes 3^-_1]_{1^-}$
two-phonon nature \cite{Brys99}.   This is a general feature in heavy
semi-magic even-even nuclei \cite{Pono98a}.  All other $1^-$
configurations have excitation energies more than 1~MeV
higher. Therefore, the $1^-_1$ state has an almost pure two-phonon
character in semi-magic nuclei.  In contrast, there are many $[qp
\otimes 1 ph]$ and $[qp \otimes 2 ph]$ configurations with the same
spin and parity close to the two corresponding configurations
$[3s_{1/2} \otimes [2^+_1 \otimes 3^-_1]_{1^-}]_{1/2^-,3/2^-}$ in
$^{117}$Sn. Interactions lead to a strong fragmentation of these two
main configurations (see, Table \ref{vt1}).  The resulting states are
carrying a fraction of the $E1$ excitation strength from the ground
state.

The predicted properties of some states with spin and parity $J^{\pi}
= 1/2^-$ and $3/2^-$ which can be excited from the $1/2^+$ ground
state in $^{117}$Sn by electric dipole transition are presented in
Table~\ref{vt1}. A large part of the $[3s_{1/2} \otimes [2^+_1 \otimes
3^-_1]_{1^-}]_{1/2^-,3/2^-}$ configurations is concentrated in the
3/2$^-$ states with an excitation energy of 3.04, 3.55 and 3.56~MeV
and in the 1/2$^-$ states at 3.00 and 3.63~MeV (see, fifth column of
this table).  These states are marked with triangles in
Fig.~\ref{fig2}b (as well as four other states with a smaller
contribution of these configurations).  The $E1$ strength distribution
among low-lying levels is even more complex because 3/2$^-$ states at
2.13, 2.33 and 3.93~MeV have a noticeable contribution from the
$3p_{3/2}$ one-quasiparticle configuration (indicated in the forth
column of Table~\ref{vt1}) with a large reduced excitation matrix
element $< 3p_{3/2} ||E1|| 3s_{1/2}>$ for which there is no analogue
in the even-even core $^{116}$Sn. Also the coupling to $[3s_{1/2}
\otimes 1^-_{\mbox{\tiny GDR}}]$, which treats the core polarization
effect, is somewhat different than in the core nucleus, because the
blocking effect plays an important role in the interaction with
other configurations (see, also Ref.~\cite{Cose95}, where only the last
type of transitions have been accounted for). 
The calculated total $B(E$1)$\uparrow$ strength 
in the energy region from 2.0 to
4.0~MeV is $7.2 \cdot 10^{-3}$~e$^2$fm$^2$.
It agrees well with the calculated $B(E1, 0^+_{g.s.} \rightarrow 
[2^+ \otimes 3^-]_{1^-}) = 8.2 \cdot 10^{-3}$~e$^2$fm$^2$ in the
neighboring $^{116}$Sn nucleus \cite{Brys99}.  

The calculations indicate that among the negative parity states in
$^{117}$Sn which are relatively strongly excited from the ground
state, a few are characterized by a visible $E1$-decay into
the low-lying $3/2^+_1$ state.  These are $3/2^-$ states at 2.33, 3.65
and 3.93~MeV and $1/2^-$ state at 3.63~MeV. The state at 2.33~MeV
decays into the $3/2^+_1$ state due to single-particle transition
with a large reduced excitation matrix element $< 2d_{3/2} ||E1||
3p_{3/2}>$. The states at higher energies decay into the $3/2^+_1$
state because of an admixture of $[3p_{1/2} \otimes [2^+_1 \otimes
3^-_1]_{1^-}]_{1/2^-,3/2^-}$ configurations in their wave functions. 

Positive parity states in $^{117}$Sn are deexciting to the 1/2$^+$ ground
state by $M1$ or $E2$ or mixed $M1/E2$ transitions. 
The predicted properties of the $1/2^+$, $3/2^+$ and $5/2^+$ states in
$^{117}$Sn are presented in Table~\ref{vt2}.
The $B(E$2)$\uparrow$ strength 
distribution is dominated by the excitation of the 3/2$^+$ state at 1.27~MeV 
and the 5/2$^+$ state at 1.49~MeV.  
The wave functions of these states carry 85\% and 
60\% of the $[3s_{1/2} \otimes 2^+_1]$ configuration, respectively.
These two states correspond with a high probability to the
experimentally observed levels at 1447 and 1510 keV.
A smaller fraction of the above mentioned configuration can be found
in the  3/2$^+$ state at 2.32~MeV (5\%) and the
5/2$^+$ state at 2.23~MeV (6\%).
The rather fragmented $E2$ strength at higher energies 
(Fig.~\ref{fig2}c) is mainly due to 
$[3s_{1/2} \otimes 2^+_{4,5}]$ configurations which are much less
collective than the first one. Fragmented $E2$ strength between 2.0 and 
4.0~MeV originating from the excitation of the $2^+_{4,5}$ phonons has
also been observed in NRF experiments on the even-mass $^{116}$Sn 
nucleus \cite{Brys00}. It could be well reproduced by theoretical
calculations (see, thin lines in Fig.~\ref{fig2}a). In the odd-mass
$^{117}$Sn nucleus the corresponding strength is even more fragmented
because of the higher density of the configurations. Nevertheless,
these $E2$ excitations at high energies contribute appreciably to the
reaction cross section, because the $E2$ photon scattering cross
section is a cubic function of the excitation energy (see, Eq.~(\ref{eq})).

The $B(M$1)$\uparrow$ strength in the calculations is concentrated
mainly above 3.5~MeV as can be seen in Fig.~\ref{fig2}d. 
The wave functions of  the 1/2$^+$ and 3/2$^+$
states at these energies are very complex. The main configurations, 
responsible for the $M1$ strength, are the $[2d_{5/2,3/2} \otimes 2^+_i]$
ones which are excited because of the internal fermion structure of the
phonons (similar to $E1$ 
 $0^+_{g.s.} \to [2^+_1 \otimes 3^-_1]_{1^-}$ excitations).
They have no analogous transitions in even-even nuclei.
The configuration $[3s_{1/2} \otimes 1^+_1]$ has an excitation energy 
of about 4.2~MeV but its contribution to the structure of states below 
4~MeV is rather weak. Most of the states with the largest
$B(M$1)$\uparrow$ values have $J^{\pi} = 1/2^+$ (see, Table~\ref{vt2}).

The QPM calculations show that the two-phonon $B(E1)$ strength from
the even-even nuclei is fragmented over several states.  Even with the 
present sensitivity of our NRF-setup, it is impossible to resolve all
of these details.  Nevertheless, when all experimentally observed
transitions between 2.7 and 3.6~MeV are considered to be $E1$
transitions, the total summed $B(E1$)$\uparrow$ strength amounts to
5.93~(75)~$10^{-3} \cdot e^{2}fm^{2}$ or  91(9)--82(10)\% of
the two-phonon $B(E1)$ strength in the neighbouring nuclei $^{116}$Sn and
$^{118}$Sn. This value  is considerably higher than in
the case of $^{139}$La and $^{141}$Pr \cite{Herz95c} where less than
40\% was observed. It shows that the $1/2^{+}$ ground state spin of
$^{117}$Sn limits the possible fragmentation and hence a larger amount 
of the particle two-phonon coupled $B(E1)$ strength could be resolved
in this NRF-experiment. 

\section{Conclusions}
Nuclear resonance fluorescence experiments performed on the odd-mass
spherical nucleus $^{117}$Sn revealed a large fragmentation of the
electromagnetic strength below an excitation energy of 4~MeV. The
search for the 
fragments of the $3s_{1/2} \otimes [2^{+}_{1} \otimes 3^{-}_{1}]$
multiplet carrying the $B(E1)$ strength of the adjacent even-even
nuclei is complicated by the limited spin information.  QPM
calculations carried out for the first time in a complete
configuration space can explain the fragmentation of the excitation
strength and shed light on how the $B(E1)$, $B(M1)$ and $B(E2)$
strength is distributed over this energy region. 

\section*{Acknowledgements}

This work is part of the Research program of the Fund for Scientific
Research Flanders.  The support by the Deutsche Forschungsgemeinschaft 
(DFG) under contracts Kn 154-30 and Br 799/9-1 is gratefully
acknowledged.  V. Yu. P. acknowledges a financial support form the
Research Council of the University of Gent and NATO.


\newpage
\begin{table}
\caption{Properties of the observed levels in $^{117}$Sn.}
\label{sn117results}
\begin{tabular}{cccc}
Energy & I$_{S}$ & $g \cdot \frac{\Gamma^{2}_{0}}{\Gamma}$ &
$B(E$1)$\uparrow$ \tablenotemark[1]\\
(keV) & (eVb) & (meV) & (10$^{-3}$ $e^{2}fm^{2}$) \\ \hline
4043.6 (7) & 3.94 (113) & 16.78 (482) & 0.242 (70) \\
4027.8 (4) & 6.56 (129) & 27.70 (544) & 0.405 (80) \\
4013.6 (6) & 2.54 (77) & 10.65 (324) & 0.157 (48) \\
3994.0 (6) & 1.73 (46) & 7.17 (189) & 0.108 (29) \\
3980.9 (5) & 3.47 (68) & 14.31 (280) & 0.217 (43) \\
3949.8 (16) & 3.21 (137) & 13.03 (558) & 0.202 (87) \\
3930.4 (5) & 1.12 (29) & 4.51 (117) & 0.071 (19) \\
3920.1 (7) & 1.45 (40) & 5.79 (158) & 0.092 (25) \\
3900.2 (6  & 1.09 (29) & 4.33 (117) & 0.070 (19) \\
3883.2 (4) & 3.58 (53) & 14.06 (207) & 0.229 (34) \\
3871.3 (4) & 5.05 (65) & 19.71 (255) & 0.325 (42) \\
3788.3 (7) & 1.57 (36) & 5.87 (133) & 0.103 (24) \\
3773.3 (13) & 0.91 (39) & 3.37 (144) & 0.060 (26) \\
3761.4 (8) \tablenotemark[2] & 0.90 (32) & 3.32 (117) & 0.060 (21) \\
3749.4 (4) & 2.08 (34) & 7.62 (123) & 0.138 (22) \\
3719.8 (7) \tablenotemark[3] & 3.16 (45) & 11.38 (163) & 0.211
(30) \\  
3560.5 (6) & 0.54 (16) & 1.79 (53) & 0.038 (11) \\
3520.4 (7) & 0.53 (20) & 1.70 (64) & 0.037 (14) \\
3489.6 (3) & 5.46 (48) & 17.31 (151) & 0.389 (34) \\
3468.8 (6) & 0.47 (16) & 1.47 (49) & 0.034 (11) \\
3425.8 (9) & 0.60 (25) & 1.84 (75) & 0.044 (18) \\
3408.5 (9) & 0.52 (19) & 1.57 (58) & 0.038 (14) \\
3385.4 (4) & 1.39 (21) & 4.15 (63) & 0.102 (16) \\
3360.1 (8) & 0.55 (20) & 1.60 (59) & 0.040 (15) \\
3349.9 (3) & 3.25 (31) & 9.50 (90) & 0.242 (23) \\
3286.0 (4) & 3.57 (35) & 10.05 (97) & 0.284 (26) \\
3228.2 (7) & 12.80 (91) & 34.71 (247) & 0.986 (71) \\
3224.6 (11) & 5.71 (51) & 15.45 (137) & 0.440 (39) \\
3169.1 (4) & 3.36 (32) & 8.78 (83) & 0.264 (25) \\
3144.9 (5) & 1.11 (18) & 2.86 (45) & 0.088 (14) \\
3134.3 (6) & 0.96 (17) & 2.46 (43) & 0.076 (13) \\
3127.8 (4) \tablenotemark[2] & 1.75 (21) & 4.45 (53) & 0.139 (17) \\
3108.2 (7) & 0.83 (17) & 2.08 (43) & 0.066 (14) \\
3100.8 (7) & 0.76 (15) & 1.91 (38) & 0.061 (12) \\
3065.7 (5) \tablenotemark[2] & 1.53 (22) & 3.74 (55) & 0.124 (18) \\
2995.7 (3) & 5.48 (41) & 12.80 (96) & 0.455 (34) \\
2986.7 (3) & 7.28 (85) & 16.89 (197) & 0.606 (71) \\
2961.9 (4) & 2.66 (28) & 6.08 (63) & 0.224 (23) \\
2908.5 (4) & 2.15 (28) & 4.73 (62) & 0.184 (24) \\
2879.8 (9) & 0.58 (20) & 1.25 (42) & 0.050 (17) \\
2864.1 (11) & 0.57 (21) & 1.21 (45) & 0.049 (18) \\
2803.4 (5) \tablenotemark[2] & 1.19 (20) & 2.43 (42) & 0.105 (18) \\
2775.2 (4) & 1.02 (19) & 2.04 (37) & 0.091 (17) \\
2718.2 (4) & 1.74 (43) & 3.34 (82) & 0.159 (39) \\
2709.1 (5) & 1.46 (22) & 2.80 (42) & 0.134 (20) \\
2590.2 (5) & 1.00 (23) & 1.75 (40) & 0.096 (22) \\
2515.8 (5) & 0.72 (17) & 1.18 (28) & 0.071 (17) \\
2415.9 (3) & 1.86 (23) & 2.82 (35) & 0.191 (24) \\
2367.3 (2) & 7.86 (55) & 11.46 (80) & 0.825 (58) \\
2356.7 (8) \tablenotemark[2] & 0.80 (25) & 1.15 (36) & 0.084 (27) \\
2304.6 (5) & 0.73 (18) & 1.01 (25) & 0.079 (20) \\
2280.4 (6) & 0.45 (16) & 0.61 (22) & 0.049 (18) \\
2128.6 (4) & 1.05 (22) & 1.24 (25) & 0.123 (25) \\
2048.2 (3) & 6.20 (49) & 6.77 (54) & 0.752 (60) \\
1510.1 (3) & 4.12 (48) & 2.45 (29) & 0.679 (80) \\
1447.2 (4) & 2.31 (40) & 1.26 (22) & 0.398 (68) \\
\end{tabular}
\tablenotemark[1]{Assuming electric dipole excitations.}
\tablenotemark[2]{The $\gamma$ transition might be due to an inelastic 
decay of a higher-lying level; see Table \ref{sn117inelas}.}
\tablenotemark[3]{multiplet}
\end{table}
\newpage
\begin{table}
\caption{Suggested spin assignments for some levels in $^{117}$Sn.}
\label{sn117spin}
\begin{tabular}{cccc}
Energy (keV) & 
$\frac{W(90^{\circ})}{W(127^{\circ})}$ &
$\frac{W(90^{\circ})}{W(150^{\circ})}$ &
$J$ 
\\ \hline
3871.3 & 0.899 (146) & 0.969 (159) & $\frac{1}{2}$ $(\frac{3}{2})$ \\
3489.6 & 0.827 (89) & 0.697 (75) & $\frac{3}{2}$ \\
3349.9 & 1.020 (134) & 0.992 (139) & $\frac{1}{2}$ $(\frac{3}{2})$ \\
3286.0 & 1.139 (130) & 1.055 (135) & $\frac{1}{2}$ $(\frac{3}{2})$ \\
3228.2 & 0.825 (60) & 0.842 (58) & $\frac{3}{2}$ \\
3224.6 & 0.984 (126) & 0.939 (104) & $\frac{1}{2}$ ($\frac{3}{2})$ \\
3169.1 & 0.942 (106) & 0.670 (77) & $\frac{3}{2}$ \\
2995.7 & 0.882 (70) & 0.773 (60) & $\frac{3}{2}$ \\
2986.7 & 0.983 (174) & 0.902 (120) & $\frac{1}{2}$ $(\frac{3}{2})$ \\
2367.3 & 0.780 (53) & 0.716 (47) & $\frac{3}{2}$ \\
2048.2 & 0.952 (86) & 1.043 (95) & $\frac{1}{2}$ ($\frac{3}{2})$ \\
\end{tabular}
\end{table}
\newpage
\begin{table}
\caption{Probable inelastic transitions to the 158.562
keV ($3/2^{+}$) level in $^{117}$Sn.  The given
$B(E$1)$\uparrow$ strengths are corrected for the possible inelastic
transitions.}
\label{sn117inelas}
\begin{tabular}{ddddd}
E$_{x}$ & E$_{\gamma}$ & I$_{\gamma}$ & $g \cdot \Gamma_{0}$  & 
$B(E$1)$\uparrow$ \\
(keV) & (keV) & & (meV) & (10$^{-3}$ e$^{2} fm^{2}$) \\ \hline
3920.1 (7) & 3920.1 (7) & 100.0 & 11.4 (38) & 0.180 (60) \\
           & 3761.4 (8) & 96.0 (375) & &       \\
3286.0 (4) & 3286.0 (4) & 100.0 & 16.3 (16) & 0.439 (43) \\
           & 3127.8 (4) & 54.5 (53)    & &       \\
3224.6 (11) & 3224.6 (11) & 100.0 & 20.3 (6) & 0.577 (17) \\
            & 3065.7 (5) & 31.2 (37) & & \\
2961.9 (4) & 2961.9 (4) & 100.0 & 9.2 (11) & 0.338 (39) \\
           & 2803.4 (5) & 51.3 (78)    & &       \\
2515.8 (5) & 2515.8 (5) & 100.0 & 2.7 (8) & 0.162 (49) \\
           & 2356.7 (8) & 127.6 (435) & &       \\
\end{tabular}
\end{table}

\begin{table}[th]
\caption{Theoretical excitation energies ($E_x$) and 
$B(E1)$$\downarrow$ reduced transition probabilities
for decays into the $1/2^+_1$ ground state  and the low-lying 
$3/2^+_1$ state for negative parity states in
$^{117}$Sn. Only the states with 
$B(E1, J^{\pi} \to 1/2^+_1) > 10^{-5}$ e$^2$fm$^2$
are presented. In the last two columns a contribution of the 
quasiparticle ($\alpha^+_{J^{\pi}}$) and the 
$[3s_{1/2} \otimes [2^+_1 \otimes 3^-_1]_{1^-}]_{J^{\pi}}$
configurations to the wave functions Eq.~\protect{(\ref{wf})}
of these states is provided when it is larger than 0.1\%.
}
\label{vt1}
\begin{tabular}{ccccc}
 $E_x$ & \multicolumn{2}{c}{$B(E1)$$\downarrow$ ($10^{-3}$ e$^2$fm$^2$})  &
$\alpha^+_{J^{\pi}}$ &
$[\frac{1}{2}^+ \otimes [2^+_1 \otimes 3^-_1]_{1^-}]$ \\
 (MeV)  & $J^{\pi} \to 1/2^+_1$ &
$J^{\pi} \to 3/2^+_1$ &  & \\
\hline
\\[-3mm]
\multicolumn{5}{c}{$\underline{J^{\pi} = 3/2^-}$} \\[1mm]
 2.13 & 0.329 & 0.092 & 1.0\% &        \\
 2.33 & 0.560 & 0.180 & 1.8\% &        \\
 2.64 & 0.030 & 0.024 &       &        \\
 3.04 & 0.058 & 0.032 &       & 10.7\% \\
 3.37 & 0.121 & 0.039 & 0.4\% &        \\
 3.46 & 0.072 & 0.002 & 0.2\% &        \\ 
 3.49 & 0.011 & 0.002 &       &        \\
 3.55 & 0.551 & 0.012 &       & 47.8\% \\
 3.56 & 0.660 & 0.038 & 0.1\% & 32.0\% \\
 3.65 & 0.160 & 0.336 & 0.5\% & ~0.1\% \\
 3.75 & 0.015 & 0.003 &       &        \\
 3.78 & 0.081 & 0.033 & 0.3\% &        \\
 3.85 & 0.042 & 0.025 &       & ~2.8\% \\ 
 3.93 & 0.235 & 0.205 & 0.9\% &        \\
 4.21 & 0.087 & 0.047 & 0.3\% &        \\
 4.32 & 0.133 & 0.074 & 0.5\% &        \\[1mm]
\multicolumn{5}{c}{$\underline{J^{\pi} = 1/2^-}$} \\[1mm]
 2.95 & 0.022 & 0.023 &       & ~0.8\% \\
 3.00 & 0.698 & 0.072 & 0.1\% & 21.9\% \\
 3.63 & 0.560 & 0.250 &       & 62.9\% \\
 3.72 & 0.070 & 1.790 &       & ~4.0\%  \\ 
 4.45 & 0.020 & 0.005 &       &        \\
\end{tabular}
\end{table}

\begin{table}[th]
\caption{Theoretical excitation energies ($E_x$) and 
$B(M1)$$\downarrow$ and $B(E2)$$\downarrow$ reduced transition
probabilities for decays into the $1/2^+_1$ ground state and the
low-lying $3/2^+_1$ states of positive parity states in
$^{117}$Sn. Only the states with $B(M1, J^{\pi} \to 1/2^+_1) >
10^{-2}$ $\mu_N^2$ or $B(E2, J^{\pi} \to 1/2^+_1) > $ 1.e$^2$fm$^4$ 
are presented. 
}
\label{vt2}
\begin{tabular}{cccrr}
$E_x$ & \multicolumn{2}{c}{$B(M1)$$\downarrow$ ($\mu_N^2$)}  &
 \multicolumn{2}{c}{$B(E2)$$\downarrow$  (e$^2$fm$^4$)} \\
MeV  & $J^{\pi} \to 1/2^+_1$ & $J^{\pi} \to 3/2^+_1$ &   
       $J^{\pi} \to 1/2^+_1$ & $J^{\pi} \to 3/2^+_1$ \\
\hline
\\[-3mm]
\multicolumn{5}{c}{$\underline{J^{\pi} = 1/2^+}$} \\[1mm]
  2.14 & 0.011 & 0.001 & & \\
  3.58 & 0.039 & 0.006 & & \\
  3.66 & 0.010 & 0.014 & & \\
  3.87 & 0.012 & 0.002 & & \\
  3.89 & 0.035 & 0.012 & & \\
  4.02 & 0.045 & 0.031 & & \\
  4.10 & 0.200 &       & & \\
  4.25 & 0.027 &       & & \\[1mm]
\multicolumn{5}{c}{$\underline{J^{\pi} = 3/2^+}$} \\[1mm]
  1.27 &       &       &  353. &    1. \\
  1.39 &       &       &    1. &  353. \\
  1.78 &       & 0.005 &    2. &    7. \\
  2.32 & 0.003 &       &   20. &   12. \\
  2.47 &       &       &    2. &   10. \\
  3.07 &       &       &   10. &       \\
  3.28 &       &       &   16. &       \\
  3.56 & 0.018 & 0.009 &       &    1. \\
  3.63 & 0.010 & 0.040 &       &       \\
  3.66 & 0.001 & 0.004 &    5. &       \\
  3.87 &       & 0.001 &    8. &    2. \\
  3.88 & 0.002 &       &    4. &    1. \\
  4.05 & 0.002 &       &   10. &       \\
  4.17 &       &       &    1. &       \\
  4.42 & 0.011 & 0.004 &       &       \\[1mm]
\multicolumn{5}{c}{$\underline{J^{\pi} = 5/2^+}$} \\[1mm]
  1.01 & &  &   82. &   11. \\
  1.32 & &  &   37. &  309. \\
  1.49 & &  &  239. &   25. \\
  2.22 & &  &    1. &    4. \\
  2.23 & &  &   22. &    1. \\
  3.07 & &  &    2. &    1. \\
  3.27 & &  &   20. &       \\
  3.79 & &  &    2. &       \\
  3.87 & &  &    4. &       \\
  3.89 & &  &    8. &       \\
  4.06 & &  &    9. &       \\
  4.17 & &  &    1. &       \\
\end{tabular}
\end{table}

\begin{figure}
\caption{Observed angular distribution ratios
$W(90^{\circ})/W(150^{\circ})$ versus $W(90^{\circ})/(127^{\circ})$
(points with error bars).  The square represents the unique location
for 1/2 -- 1/2 -- 1/2 spin sequences.  Pure $E1$, $M1$ or $E2$
transitions in a 1/2 -- 3/2 -- 1/2 spin sequence are marked with the
triangle.  Mixed $E1$/$M2$ and $M1$/$E2$ transitions ($\delta$ $\neq$
0, $\pm \infty$) in this spin sequence are located on the solid
line. The star corresponds to a pure $E2$ transition in a 1/2 -- 5/2
-- 1/2 spin sequence.} 
\label{sn117angdis}
\end{figure}
\begin{figure}
\caption{Photon scattering spectra of the odd-mass $^{117}$Sn (summed
spectrum) sandwiched between those of the even-even $^{116}$Sn and
$^{118}$Sn nuclei (at a scattering angle of 127$^{\circ}$), all taken
with an endpoint energy of 4.1~MeV.} 
\label{sn117spec}
\end{figure}
\begin{figure}
\caption{Schematic level schemes for the odd-mass $^{117}$Sn nucleus
and the even-even core nucleus $^{116}$Sn. The level scheme for $^{117}$Sn
can be obtained in a phenomenological simple core coupling model by
coupling the 3s$_{1/2}$ neutron to the even-even $^{116}$Sn neighbour.} 
\label{sn117scheme}
\end{figure}
\begin{figure}
\caption{Total integrated photon scattering cross sections
$I_{S}$ observed in $^{117}$Sn (b)) centered between those observed in
its even-even neighbours $^{116}$Sn (a)) and $^{118}$Sn (d))
\protect\cite{Brys00}. The integrated elastic photon scattering cross
sections calculated within the QPM are included in panel c) for
comparison. The lines marked by  a triangle correspond to
levels with a 3/2 spin.  The squares represent levels with
a 1/2 (3/2) spin.}
\label{sn117is} 
\end{figure}

\begin{figure}
\caption{Calculated integrated elastic photon scattering cross
sections $I_{S}$ in $^{116}$Sn a) and $^{117}$Sn b) -d). $I_S$ for
$E1$ decays are plotted by thick lines in a) and b). The $E1$ decays
in b) which are predominantly due to
$[3s_{1/2} \otimes [2^+_1 \otimes 3^-_1]_{1^-}]_{1/2^-,3/2^-} 
\to 3s_{1/2}$ transitions are marked by triangles.
\label{fig2} }
\end{figure}

\newpage
\begin{figure}
\epsfig{file=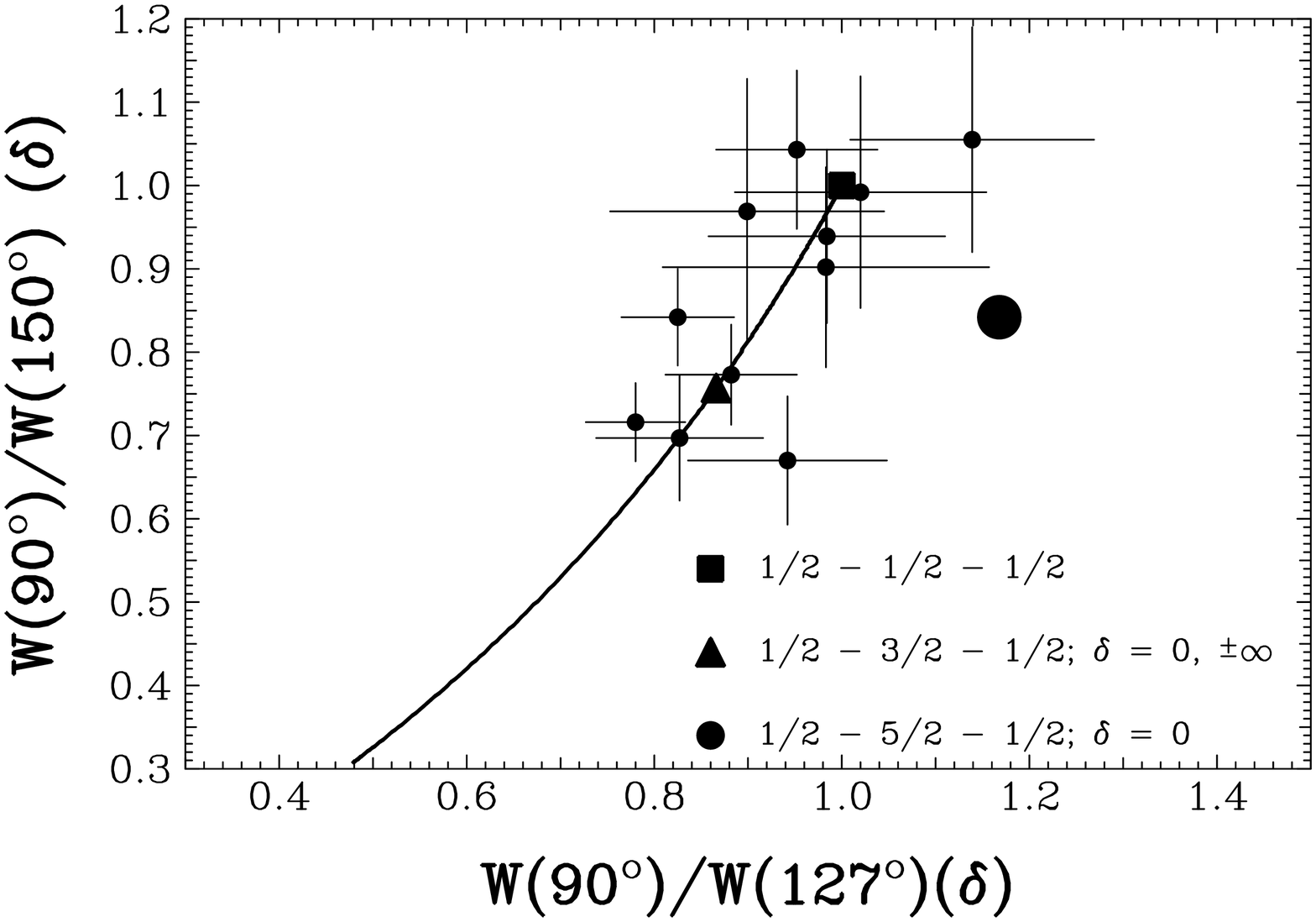,width=12cm}
\end{figure}
{\Huge Figure 1}
\begin{figure}
\epsfig{file=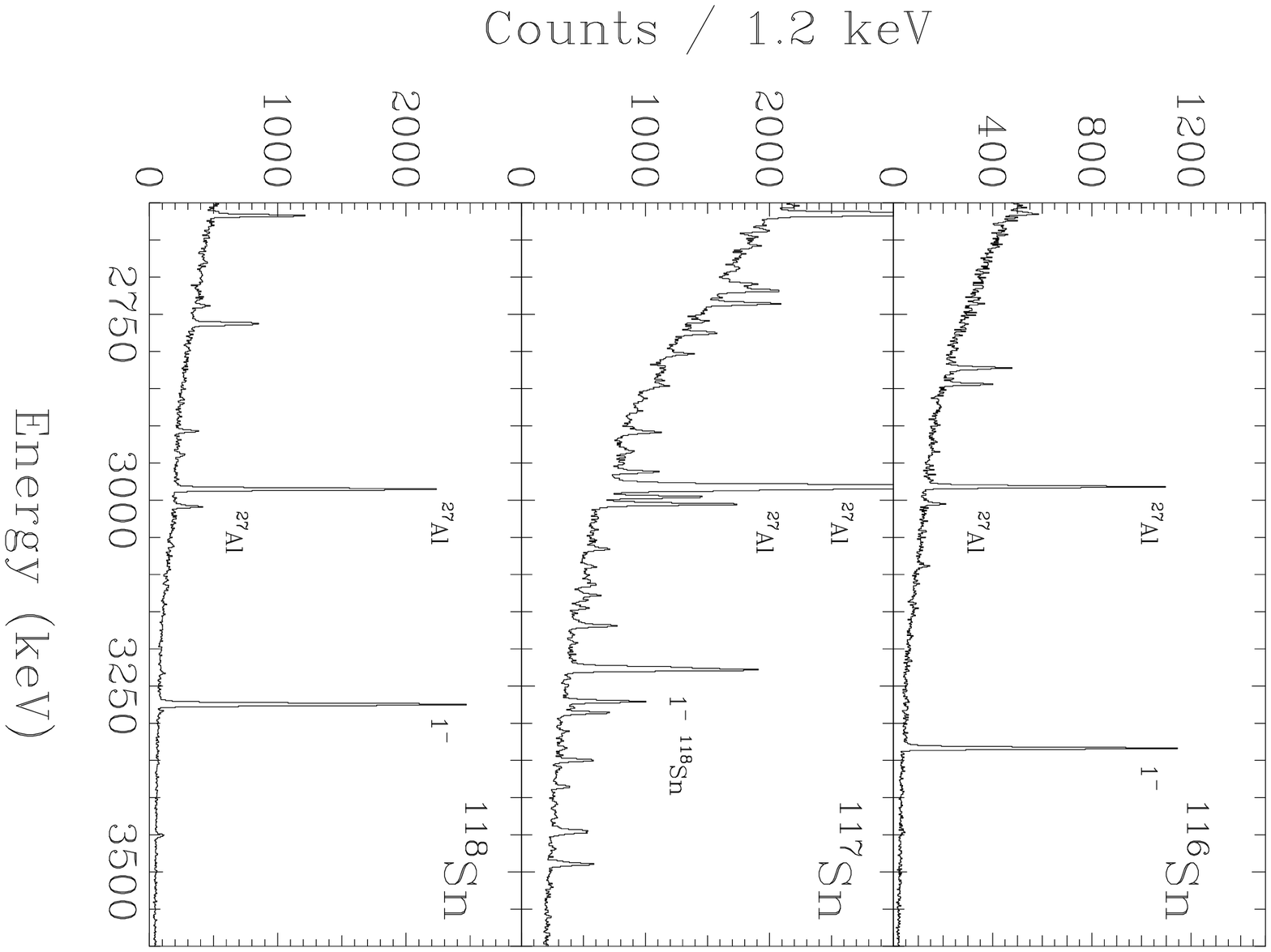,width=12cm,angle=90}
\end{figure}
{\Huge Figure 2}
\newpage
\begin{figure}
\epsfig{file=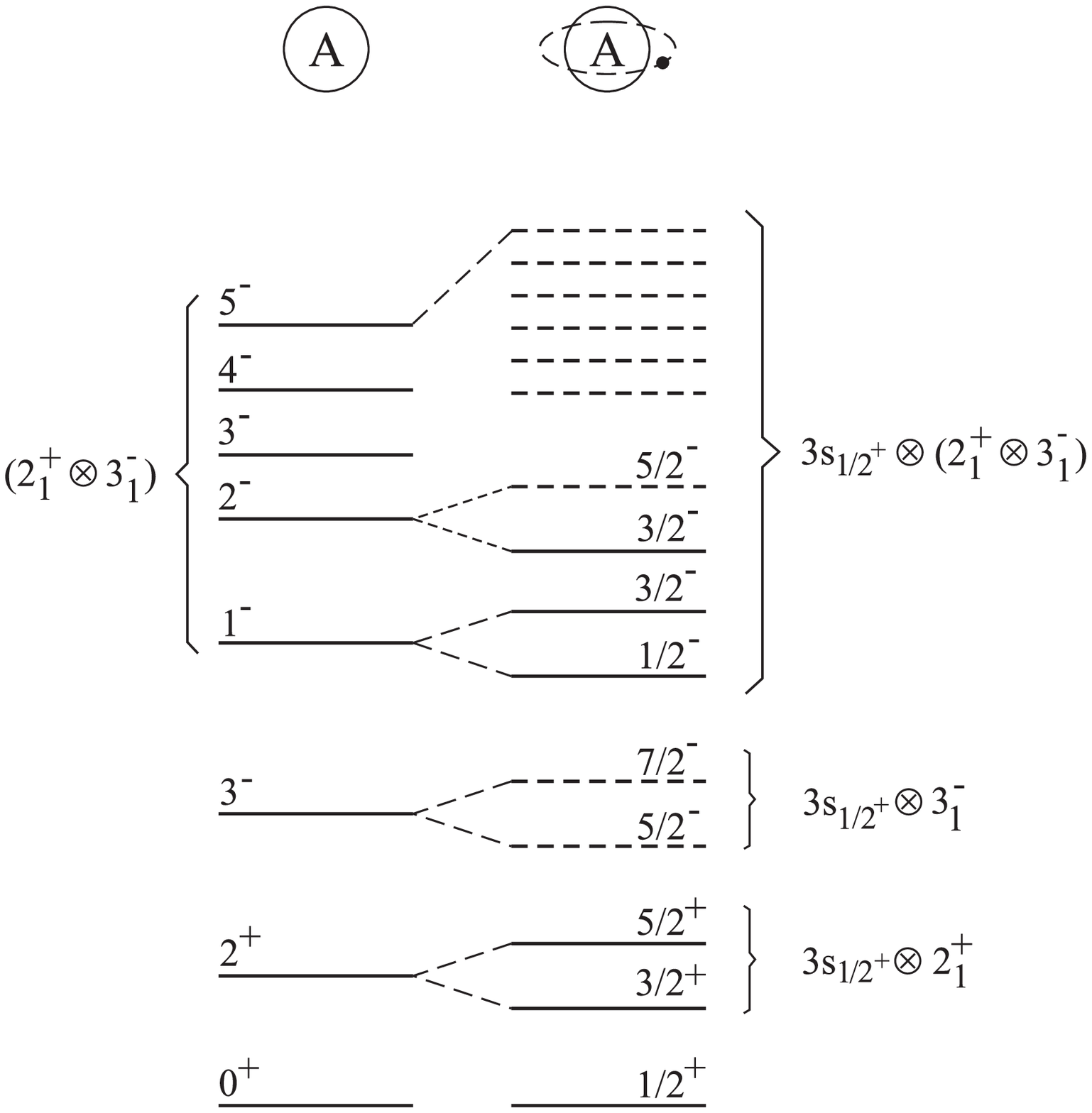,width=12cm}
\end{figure}
{\Huge Figure 3}
\newpage
\begin{figure}
\epsfig{file=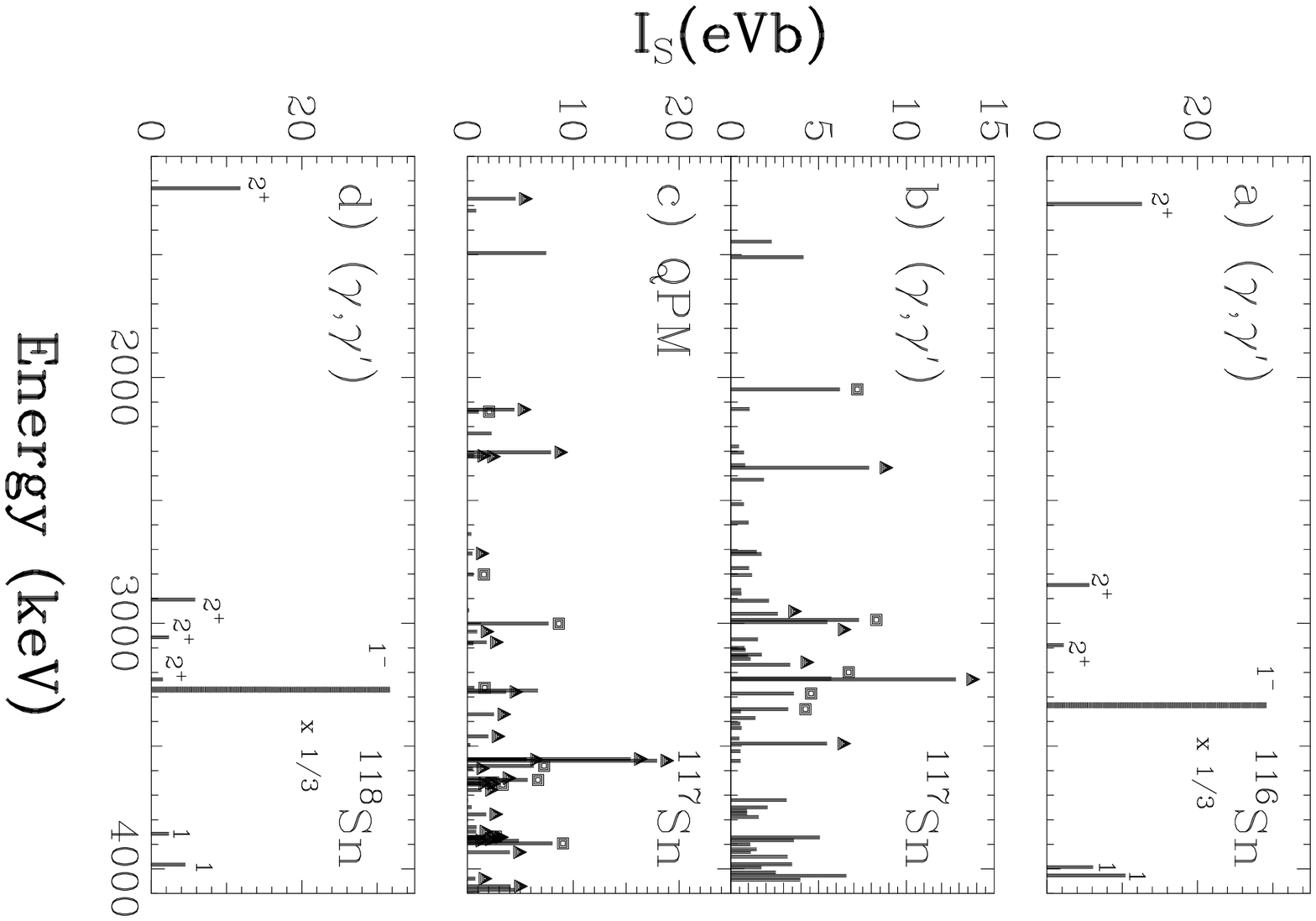,width=12cm,angle=90}
\end{figure}
{\Huge Figure 4}
\begin{figure}
\epsfig{file=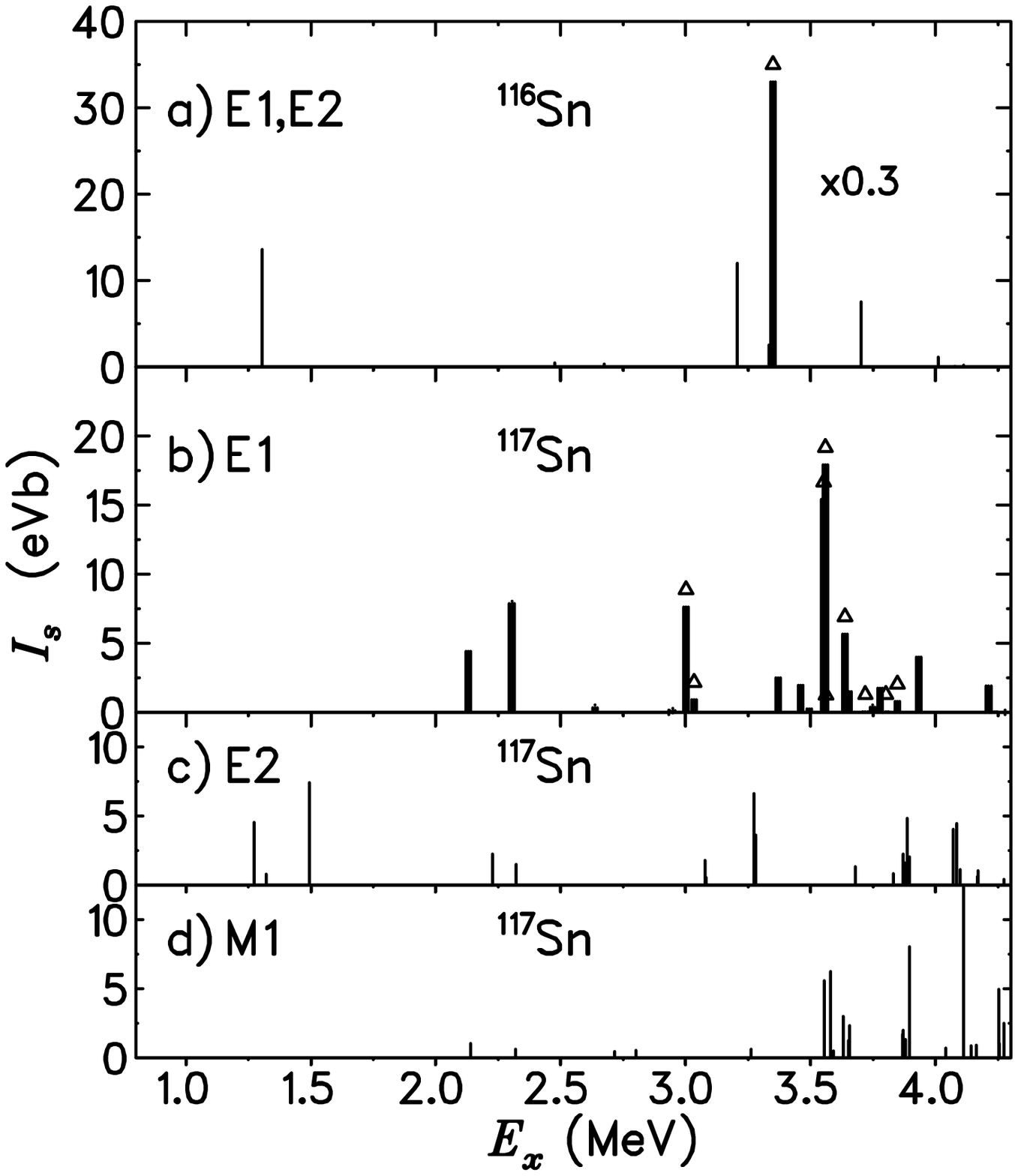,width=12cm}
\end{figure}
{\Huge Figure 5}
\newpage
\end{document}